\documentclass[twocolumn,showpacs,preprintnumbers,amsmath,amssymb]{revtex4}

\usepackage{graphicx}
\usepackage{dcolumn}
\usepackage{bm}

\begin{document}

\title{Stable Mean Field Solution of a Short-Range Interacting SO(3) Quantum Heisenberg
                        Spin-Glass \\}

\author{C.M.S. da Concei\c c\~ao$^1$ and E.C.Marino$^{1,2}$}
\affiliation{
 $^1$Departamento de F\'\i sica, Universidade Federal do Rio de Janeiro, Cx.P. 68528, Rio de Janeiro, RJ 21941-972, Brazil}
\affiliation{$^2$ Department of Physics, Princeton University, Princeton, NJ 08544, USA}


\date{\today}

\begin{abstract}

We present a mean-field solution for a quantum, short-range interacting, disordered, SO(3) Heisenberg spin model, in which the Gaussian distribution of couplings is centered in an AF coupling $\bar J>0$, and which, for weak disorder, can be treated as a perturbation of the pure AF Heisenberg system. The phase diagram contains, apart from a N\'eel phase at $T=0$, spin-glass and paramagnetic phases whose thermodynamic stability is demonstrated by an analysis of the Hessian matrix of the free-energy. The magnetic susceptibilities exhibit the typical cusp of a spin-glass transition.

\end{abstract}

\maketitle

Spin glasses (SG) have attracted a lot of interest since a long time \cite{by,mpv}. They are characterized by having some
degree of frustration, which is caused by the competition between different types of order. As a consequence
some of their properties are
shared with paramagnetic states and some other with ordered, ferromagnetic or N\'eel states.
This is usually produced by a random distribution of coupling constants that allow for interactions
of opposite signs. The
characteristic time scale of this is typically much larger than the dynamical time scale intrinsic to the system, a fact that
leads to the so-called ``quenched'' thermodynamical description \cite{by,mpv}.

In a pioneering work, Edwards and Anderson (EA) proposed a model for SG,
introduced an order parameter for the detection of a SG phase and
employed the replica
method to deal with the quenched average \cite{ea}.
A simplified version of the EA model describing classical Ising spins with long-range interactions
was solved for the first time by
Sherrington and Kirkpatrick  (SK) using the mean-field approach \cite{sk}.
The solution, however, proved to be unstable \cite{at} and this fact has been generally attributed to the so-called replica
symmetry it possesses.
Indeed, stable replica symmetry breaking solutions were subsequently found
\cite{par} and a lot of knowledge has been gathered about long-range interacting spin glasses since then \cite{by}.

 The realistic systems found in nature, however, are most likely short-range interacting, quantum SO(3) Heisenberg spin systems, which for this reason
are especially appealing from the physical point of view.
 Interestingly, however, after more than thirty years, very little is known about the properties of short-range quantum spin-glasses,
 especially with SO(3) symmetry.
Numerical calculations exist \cite{renum}, but very few
analytical approaches are available.
Interesting results, however, have been obtained in related systems: a
quantum, long-range SG model \cite{bm} has been solved recently \cite{persach}, whereas a Landau-Ginzburg, phenomenological
approach has been developed for quantum,
short-range interacting, rotor and Ising SG models \cite{rsy}.

Nevertheless, still a number of open questions concerning quantum short-range SO(3) SG systems remain:
is a mean-field approach possible; is there a SG phase; is it stable;
is it replica symmetric; is there any basic clash between replica symmetry and stability?

It is natural to expect that a continuum description would be useful for a short-range SG system, since the nearest neighbor interactions
become just derivatives in this limit. A serious obstacle arises, however when taking the continuum limit of a quantum system. This is
connected with
the quantum Berry phases, which in general would not cancel when summed over the lattice \cite{rsy,sy}.

In this work we consider a model for a disordered, short-range SO(3) quantum spin system in which we
circumvent this problem and obtain a continuum model that will enable us to address the previous questions.
This is achieved by introducing disorder as a perturbation of an antiferromagnetic (AF) 2D Heisenberg model, for which
the sum of the quantum phases is known to cancel \cite{xwef,ha1}. The situation is
completely different from the original EA model, where $\bar J =0$, and consequently the disorder cannot be taken as a perturbation of a Heisenberg system \cite{ea}.
Using the continuum description, we extract the $T\times \bar J$ phase diagram of the system from the mean field solution, which presents replica symmetry.
This exhibits a N\'eel phase at $T=0$, whose quantum critical point is displaced from its original value in the pure system. It also contains spin glass (SG) and paramagnetic (PM) phases, whose thermodynamical stability is demonstrated by a careful analysis of the Hessian matrix of the average free-energy.

There is in addition an appealing physical motivation for this model, in connection to the high-Tc cuprates.
Indeed, these materials, when undoped, are 2D Heisenberg antiferromagnets, which upon doping, develop a SG phase before becoming superconductors.
Our model, describing precisely the AF-SG transition, is therefore potentially useful for studying the magnetic fluctuations of such materials.

The model consists of
an SO(3) quantum Heisenberg-like hamiltonian,
containing only nearest neighbor interactions of the spin operators $\mathbf{\widehat{S}}_{i}$, on the sites of a $2D$ square lattice
of spacing $a$.
The couplings $J_{ij}$ are random and associated with a Gaussian probability distribution
$P[J_{ij}]$ with variance $\Delta J$ and
centered in $\bar{J}>0$, such that $\Delta J\ll\bar J$. We consider the quenched situation, in which, according to the replica method \cite{by,ea} the average free-energy is given by
$\overline{F}=-k_{B}T\lim_{n\longrightarrow
0}\frac{1}{n}([Z^{n}]_{av}-1),
$
where $Z^{n}$ is the replicated partition function for a given configuration of couplings $J_{ij}$
and $[Z^{n}]_{av}$ is the average thereof with the Gaussian distribution.

Using the coherent spin states
$
|\mathbf{\Omega}_{i}^{\alpha}(\tau)\rangle $, such that
$\langle\mathbf{\Omega}_{i}^{\alpha}(\tau)|\mathbf{\widehat{S}}^\alpha_{i}
|\mathbf{\Omega}_{i}^{\alpha}(\tau)\rangle = S \mathbf{\Omega}_{i}^{\alpha}(\tau)$
($i$: lattice sites, $\alpha$: replicas, $\tau$: euclidian time, $S$: spin quantum number)
 \cite{ha1},
we may express $Z^{n}$  as a functional integral over
the classical spin $\mathbf{\Omega}_{i}^{\alpha}(\tau)$. The average over the disordered couplings $J_{ij}$
can then be performed, yielding
\begin{equation}
[Z^{n}]_{av}=\int\mathcal{D}\mathbf{\Omega}\mathcal{D}\mathbf{Q} e^{- \int_0^\beta  L_{\bar J,\Delta}d\tau }, \label{hamiltonianoefetivo1}
\end{equation}
where
$
L_{\bar J,\Delta}= \sum_{i,\alpha} L^B_{i,\alpha}
+S^{2}\bar J \sum_{\langle
ij\rangle}\mathbf{\Omega}_{i}^{\alpha}(\tau)\cdot\mathbf{\Omega}_{j}^{\alpha}(\tau) + L_\Delta$.

In this expression, $L^B_{i,\alpha}$ are the quantum phases \cite{sach} and $L_\Delta$ is a quartic interaction term,
proportional to $(\Delta J)^{2}$, which is generated by the Gaussian average. We use the
standard Hubbard-Stratonovitch procedure in order to replace the quartic interaction by a trilinear
interaction of $\Omega_{i,a}^{\alpha}(\tau)$ with the variables $Q_{i,ab}^{\alpha\beta}(\tau,\tau')$, where $ab$ are SO(3) indices.

This is no longer a disordered system. The  disorder, which was originally present manifests now through the interaction term,
proportional to $(\Delta J)^{2}$. In the absence of disorder, we would have $\Delta J \rightarrow 0$ and $[Z^{n}]_{av} $ would reduce
to the usual coherent spin representation of the AF Heisenberg model, with a coupling $\bar J > 0$ \cite{sach,chn,em}.

Since we are only considering the weakly disordered case $(\Delta J \ll \bar J)$ our model, described by the
effective lagrangian in (\ref{hamiltonianoefetivo1})
is a perturbation of the AF 2D quantum Heisenberg model. This means we can decompose the classical spin
$\Omega_{i,a}^{\alpha}(\tau)$ into antiferromagnetic and ferromagnetic fluctuations as in that model \cite{sach}. Using this, then
it follows that the sum of the quantum Berry phases,
$ L^B_{i,\alpha}$, over all the lattice sites vanishes, as in the pure system \cite{ha1,xwef}.

We can therefore take the continuum limit in the usual way as in the pure AF 2D quantum Heisenberg model \cite{ha1,ha2,sach}
obtaining an SO(3) generalized relativistic nonlinear sigma model (NLSM). This contains the field $\vec n^\alpha=(\sigma^\alpha, \vec \pi^\alpha)$, which is
the continuum limit of the (staggered) spin $\mathbf{\Omega}^\alpha$ and satisfies the constraint
$\vec n^\alpha\cdot\vec n^\alpha= \rho_s $; where $\rho_s=S^2 \bar J$.
 The generalized NLSM also contains a trilinear interaction of $\vec n^\alpha$ with the
 Hubbard-Stratonovitch field $  Q_{ab}^{\alpha\beta}(\tau,\tau')$, which is proportional to
 $(\Delta J)^2$ and corresponds to $L_\Delta$.

Notice that a null
 value for $\bar J$, as we have in the EA model \cite{ea} would make the perturbation around a NLSM meaningless. A negative value, on
  the other hand, would correspond
 to the ferromagnetic Heisenberg model, which
 after taking the continuum limit, is associated to the
 non-relativistic NLSM. Here perturbation would be possible, however, the Berry's phases would no longer cancel. We emphasize, therefore, the enormous difference that exists, both from the physical
 and mathematical points of view, in considering $\bar J$ as positive, negative or null in the Gaussian distribution of the EA model.

 Using the decomposition ($q^{\alpha\beta}=0$ for $\alpha = \beta$)
 $  Q_{ab}^{\alpha\beta}(\vec r;\tau,\tau')\equiv \delta_{ab}[\delta^{\alpha\beta}\chi(\vec r;\tau,\tau')+
q^{\alpha\beta}(\vec r;\tau,\tau')]$ and enforcing the NLSM constraint with the lagrangian multiplier field $\lambda$, as usual,
we integrate on the $\vec \pi^\alpha$ component
of $\vec n^\alpha$, obtaining
\begin{equation}
[Z^{n}]_{av}=\int\mathcal{D}\sigma\mathcal{D}\chi\mathcal{D}q\mathcal{D}\lambda
e^{-S_{\mathrm{eff}}[\sigma,\chi,q,\lambda]} \label{gm1}.
\end{equation}

We evaluate $[Z^{n}]_{av}$ by means of a
stationary phase approximation, obtaining
\begin{equation}
[Z^{n}]_{av}= e^{- n\bar {S}_{\mathrm{eff}}\left[\sigma_{\mathrm{ex}}^\alpha, m^2, q_{\mathrm{ex}}^{\alpha\beta}(\tau-\tau'), \chi_{\mathrm{ex}}(\tau-\tau')\right]},\label{funcaoparticaosigm}
\end{equation}
where
$n \bar S_{\mathrm{eff}}$
is the effective action in (\ref{gm1}), evaluated at the extremant
configurations $\sigma^\alpha(\mathbf{r},\tau)=\sigma_{\mathrm{ex}}^\alpha$, $2i\lambda_{\mathrm{ex}}=m^2$, $\chi(\mathbf{r},\tau,\tau')= \chi_{\mathrm{ex}}(\tau-\tau')$ and
$q^{\alpha\beta}(\mathbf{r},\tau,\tau')=q_{\mathrm{ex}}^{\alpha\beta}(\tau-\tau')$
(we henceforth neglect the ``${\mathrm{ex}}$'' subscript).

The parameter $m^2$, as usual, is a spin gap
scale, such that its inverse is the correlation length. It will be determined by the temperature, $\bar J$ and $\Delta J$, as we show below.
The staggered magnetization, $\sigma$ characterizing an ordered AF state, is defined as
$\sigma^2 = \lim_{n\rightarrow 0} \frac{1}{n} \sum_{\alpha} \sigma^2_\alpha$.

By taking the limit $n\rightarrow 0$ in (\ref{funcaoparticaosigm}) we immediately realize that
the average free-energy is given by $\bar F= \frac{1}{\beta} \bar S_{\mathrm{eff}}$.
This is most conveniently expressed in
the space of Matsubara frequencies $\omega_r=2\pi rT, r \in \mathbb{Z}$. Fourier transforming  $\chi$'s and $q$'s
we obtain the average free-energy density as a functional
$
\bar f = \bar f \left[\sigma^\alpha, m^2, q^{\alpha\beta}(\omega_r), \chi(\omega_r)\right],
$
where $\chi(\omega_r)$ and $q^{\alpha\beta}(\omega_r) $ are, respectively, the Fourier components of
 $\chi(\tau-\tau')$ and
$q^{\alpha\beta}(\tau-\tau')$, for which we
use the simplified notation
$\chi_r\equiv \chi(\omega_{r})$ and $  q^{\alpha\beta}_r\equiv q^{\alpha\beta}(\omega_{r})$.

From (\ref{hamiltonianoefetivo1}), we can show that
$
Q_{i}^{\alpha\beta}(\tau,\tau') = \langle \hat S_i^\alpha(\tau) \hat S_i^\beta(\tau') \rangle
$
and therefore, according to the previous decomposition of $Q$ into $\chi$'s and $q$'s, we can identify $\chi_0 $ as the static magnetic susceptibility, whereas the integrated susceptibility is given by $\chi_\mathrm{I}= \sum_r \chi_{r} $.
The EA order parameter for the SG phase \cite{ea,by}, accordingly, is given by
 $q_{\mathrm{EA}} = T \bar q_0 $, where $\bar q_0 =\lim_{n\rightarrow 0} \frac{1}{n(n-1)} \sum_{\alpha\beta} q_0^{\alpha\beta}$.

By taking the variations of $\bar f$ with respect to the variables $\sigma^\alpha, m^2, q^{\alpha\beta}_r, \chi_r$,
we obtain the mean field equations (MFE). These possess a replica symmetric solution that will allow us to determine the phase diagram of the system.

Let us begin with the search for an ordered N\'eel phase. The MFE imply that this may only occur at $T=0$, in agreement with \cite{mw}. Indeed, we find that on the line $( T=0, \rho_s>\rho_0)$: $\sigma^2=\frac{1}{8}[\rho_s-\rho_0]$,
$ q_{\mathrm{EA}}^{\mathrm{AF}}=\frac{1}{4\rho_s}[\rho_s-\rho_0] $,
$\chi_0^{\mathrm{AF}}, \chi_{\mathrm{I}}^{\mathrm{AF}} \rightarrow  \infty$. Here
$\rho_0=\frac{\Lambda}{2\pi}\left[1+\frac{1}{\gamma}\left[1+\frac{1}{2}\ln(1+\gamma)\right]\right]$, where $\gamma=3\pi\left(\frac{ \bar J}{\Delta J}\right)^2$ and $\Lambda=1/a$. We also find
$m^2=0$, so the correlation length diverges on this line.

 The previous results characterize an AF ordered N\'eel phase
$(\sigma \neq 0, q_{\mathrm{EA}} \neq 0)$ on the line $(T=0, \rho_s>\rho_0)$. The parameter $\gamma$ appears naturally in the calculation.  $\frac{1}{\gamma}$ is a measure of the amount of frustration in the system and the actual perturbation parameter.
Since we are working in the regime of weak disorder, we take $\gamma \gg 1$. We see again that a disorder perturbation would be impossible in the original EA model, where $\gamma=0$. In the absence of disorder $(\Delta J=0, \gamma \rightarrow\infty)$,
$\rho_0\rightarrow \rho_0(0)=\frac{\Lambda}{2\pi}$, which is the well-known quantum critical coupling determining the boundary of the AF phase in the pure 2D AF Heisenberg model at $T=0$ \cite{chn,sach,em}. The effect of disorder on the AF phase is to displace the quantum critical point (QCP) to the right.
This result should be expected on physical grounds: in the presence of disorder a larger coupling is required, to stabilize an
ordered AF phase.

We now search for PM and SG phases. In both of them we have $\sigma =0$.
From the MFE we may determine $\chi_r$ and $q_r$. We find, in particular,
$\bar q_0 = 0$, for $m^2 > m_0^2$ and $\bar q_0 =(3/A)[m_0^2-m^2] >0$, for $m^2 < m_0^2$. In these expressions, $m_0^2=\frac{\Lambda^2}{\gamma}[1+\ln(1+\gamma)]$ and $A=\gamma/(6\pi\rho_s\Lambda^2)$.
Since $q_{\mathrm{EA}} = T \bar q_0$  it follows that $\bar q_0$ is also a SG order parameter and
we conclude that the former phase ($m^2 > m_0^2$) is paramagnetic $(\sigma = 0, q_{\mathrm{EA}} = 0)$, whereas the latter ($m^2 < m_0^2$) is a SG phase $(\sigma = 0, q_{\mathrm{EA}} \neq 0)$.
The phase transition occurs at $m^2 = m_0^2$. In the unperturbed limit where the disorder is removed $(\gamma \rightarrow\infty)$, we would have $m_0^2=0$ and the SG phase would no longer exist.

We can determine the susceptibilities from the MFE. In the PM phase ($m^2 > m_0^2$), we get
$\chi_\mathrm{I}^{\mathrm{PM}}= \frac{1}{3T} $ and
$
\chi_0^{\mathrm{PM}}= \frac{1}{3T}- \mathrm{Y}(m^2,T,\rho_s,\gamma)\equiv X(m^2,T,\rho_s,\gamma)$, where
the function $ \mathrm{Y}(m^2,T,\rho_s,\gamma) $ is explicitly obtained.
It has the properties $\mathrm{Y}\stackrel{T>>\Lambda}\longrightarrow 0$
and $\mathrm{Y}\stackrel{T\rightarrow 0}\longrightarrow \frac{1}{3T}[\rho_0/\rho_s]$ and
 also $\mathrm{X}\stackrel{m^2 \rightarrow m_0^2}\longrightarrow \bar\chi_{\mathrm{cr}} = \frac{1}{6\pi \rho_s}\ln(1+\gamma)$, implying that
 the critical value of  $\chi_0^{\mathrm{PM}}$ is
  $\bar\chi_{\mathrm{cr}} $.

We see that $\chi_0^{\mathrm{PM}}$ satisfies the Curie law at high-temperatures and diverges as
$\chi_0^{\mathrm{PM}} \stackrel{T\rightarrow 0}\longrightarrow \frac{1}{3T}[1-\rho_0/\rho_s]$, for $T\rightarrow 0$. This is the expected behavior
for $(\rho_s >\rho_0)$, where the AF phase appears. For $(\rho_s <\rho_0)$, conversely, we will see that the PM-SG phase transition occurs at a finite $T_c$ (Fig.1)
and the previous expression is no longer valid.

We now turn to the the SG phase ($m^2 < m_0^2$).
Using the MFE we obtain
$
\chi_0^{\mathrm{SG}}= X -
A\Delta m_0
$
and
$
\chi_\mathrm{I}^{\mathrm{SG}}=\frac{1}{3T}-A\Delta m_0,
 $
where $\Delta m_0 \equiv [m_0^2-m^2]$ .

We can identify a clear cusp at the transition both in the integrated and static susceptibilities.
This is an important result, since
the presence of
these cusps is a benchmark of the SG transition and
has been experimentally observed in many materials presenting a SG phase \cite{by}.

We may determine the critical curve $T_c \times \rho_s$ by
observing that the critical condition $m^2=m_0^2$ implies
$\bar\chi_{\mathrm{cr}}-X(m_0^2,T_c,\rho_s,\gamma) =0$.
For $T_c\ll\Lambda$, which corresponds the situation found in realistic systems, this becomes, near the
quantum critical point ($\rho_s \lesssim \rho_0$)
\begin{equation}
\frac{T_c}{2\pi}\left[ \ln\left( \frac{\Lambda}{T_c}\right)^2-\ln(1+\gamma)\right]=\rho_0-\rho_s\label{ccritica}.
\end{equation}
We plot the corresponding phase diagram in Fig.1.
\begin{figure}[ht]
\centerline {
\includegraphics
[clip,width=0.5\textwidth
,angle=0
] {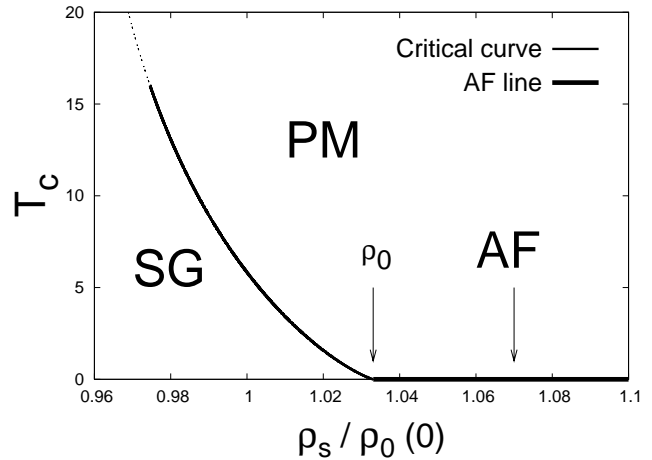} }
\caption{Phase diagram for $\gamma=10^2,\Lambda=10^3$. The critical curve corresponds to (\ref{ccritica}) and is valid near the QCP $\rho_0$ (solid curve). $\rho_0(0)=\Lambda/2\pi$ is the QCP of the pure AF system. Notice that disorder besides creating the SG phase, displaces the QCP to the right. The value ascribed to $\Lambda$ is a realistic one in $K$ ($\Lambda \rightarrow \frac{\hbar v_s}{k_{B}}\Lambda$; $v_s$: spin-wave velocity). The resulting temperatures naturally appear with the correct
order of magnitude, in $K$, found in real SG systems \cite{by}.}
\label{FigRhoxTc}
\end{figure}

The critical behavior of relevant quantities may be determined by analyzing the function $Y(m_0^2,T,\rho_s,\gamma)$ for $T\sim T_c$
and $m^2 \sim m_0^2$.
This yields, near the transition, for $\rho_s < \rho_0$,
\begin{equation}
X \sim \left(\frac{T_c}{T}\right) \bar\chi_{\mathrm{cr}}\ ;\ [ m^2-m_0^2] \sim 4\pi\Lambda \left[\frac{T-T_c}{T_c}\right][\rho_0-\rho_s].
\label{chipm}
\end{equation}
From this we can fully determine the critical behavior
of the SG order parameter and susceptibilities (see Fig.2).
\begin{figure}[ht]
\centerline {
\includegraphics
[clip,width=0.5\textwidth
,angle=0
] {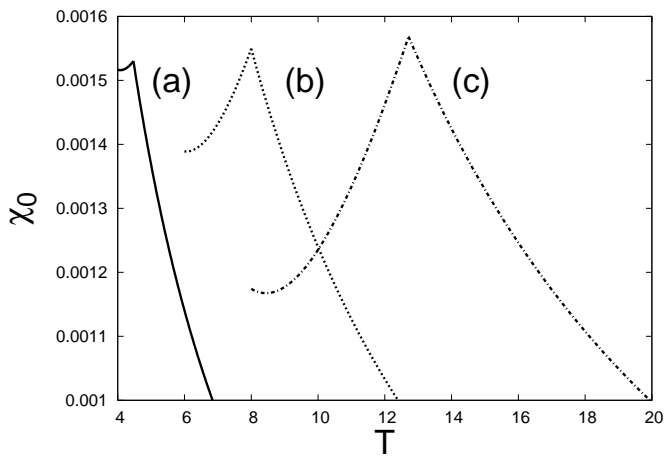} }
\caption{The static susceptibility for different values of $\rho_s$:
$\rho_s/\rho_0(0)= $(a)1.005 ,(b)0.993, (c)0.981.
 The cusps, characteristic of the SG transition, occur at the corresponding critical temperatures
(in $K$). $\chi_0$ is in $K^{-1}$.}
\label{Chi0-1}
\end{figure}
For $\rho_s>\rho_0$ and $T>0$, we always have $m^2>m_0^2$ and $\sigma=\bar q_0=0$ i.e. the system is in the PM phase for any
finite temperature.
A quantum phase transition occurs at the point ($T=0, \rho_s=\rho_0$), connecting the SG phase to the AF phase.

Let us now examine the $\Delta J$ dependence of the phase diagram. For this we make $\Delta J\rightarrow \Delta J(1+ \epsilon)$, with $|\epsilon| \ll 1$, for fixed $\rho_s$, and study how the
relevant quantities change. We find, for $\rho_s<\rho_0$,
\begin{equation}
\frac{\Delta\rho^\epsilon-\Delta\rho}{\Delta\rho} = \frac{T_c^\epsilon-T_c}{T_c}=
\frac{\Delta m_0^\epsilon-\Delta m_0}{4\pi\Lambda\rho_0}= \epsilon\frac{\Lambda}{2\pi\gamma\Delta\rho}\ln\gamma
\label{depdes},
\end{equation}
where $\Delta\rho=\rho_0-\rho_s$.
Also, $[\bar q_0^{\mathrm{SG}}]^\epsilon-\bar q_0^{\mathrm{SG}}= \frac{\epsilon}{\pi\rho_s}\ln\gamma $. We see that increasing the amount of disorder, through an increment of the Gaussian width,
will increase $\rho_0$ and, for a fixed $\rho_s$, also $T_c$, $m_0^2$ and $\bar q_0^{\mathrm{SG}}$. Conversely, decreasing the amount of disorder by narrowing the Gaussian width will produce the opposite effects.

We finally consider the question of the thermodynamic stability of the phases. We will focus on the SG and PM phases.
The stability of the AF phase should not be a problem and will be considered elsewhere, in an extended version of this paper. The stability of the SG phase is the main concern here.
For investigating this point, we have determined the Hessian matrix of the free-energy
density $ \bar f[\sigma^\alpha; q^{\alpha\beta}(\omega_0),...q^{\alpha\beta}(\omega_r)...; m^2;\chi(\omega_0),..., \chi(\omega_r),...] $.

This is a matrix with entries of dimensions $\lim_{n\rightarrow 0}[n;n(n-1)_0,...,n(n-1)_r,...;1;1_0,...,1_r,...]$ corresponding, respectively,
to derivatives with respect to each of the above variables. A sufficient condition for the mean field solution to be a local minimum is to have all
the principal minors of the Hessian positive. This would rule out the usual instabilities found in long-range replica-symmetric solutions, but of course, not meta-stability. This would deserve further investigation.

 We have carefully evaluated each of these determinants
in the limit $n \rightarrow 0$, for $\sigma =0$ (PM and SG phases).
We obtain $D_\sigma=1$. All the remaining principal minors, namely, $D_{q_0}$,..., $D_{q_r}$,..., $D_{m^2}$, $D_{\chi_0}$,..., $D_{\chi_r}$,...
can be written in the form
\begin{equation}
\xi (\gamma-G_0) + \eta \bar q_0
\label{pm},
\end{equation}
where $\xi$ and $\eta$ are positive real numbers (in the case of $D_{q_0}$, for instance, we have $\xi=\eta=1/\gamma$) and $G_0 $
is such that in the PM phase $G_0<\gamma$ and in the SG phase $G_0 =\gamma$.

In the PM phase we have $\bar q_0 =0$ and $G_0 < \gamma$, therefore it follows that
all the principal minors are positive. In the SG phase, conversely, $(\gamma-G_0)=0$ and
$\bar q_0 > 0$ and we conclude that also in the SG phase all the principal minors are positive. At the transition, all of
them, except $D_\sigma$, vanish, since both $(\gamma-G_0)=0$ and $\bar q_0 =0$.
The above result establishes the thermodynamic stability of the SG and PM phases presented above.

Our results, based on a mean-field approach to a short-range interacting, weakly disordered, SO(3) quantum spin system, clearly show the existence of a stable SG phase, at a finite $T$. The behavior of the susceptibilities, exhibiting the characteristic cusps at the
transition is an evident manifestation of it.
The solution has all $q^{\alpha\beta}$ equal, being therefore replica symmetric.
The fact that it is stable
seems to indicate that,
in the case of short-ranged interactions,
there is no basic clash between replica symmetry and the stability of the mean-field solution.

ECM would like to thank Curt Callan and the Physics Department of Princeton University for the kind hospitality.
This work was supported in part by CNPq and FAPERJ. CMSC was supported by CAPES. We are grateful to P.R.Wells for the help
with the graphics.

\bibliography{apssamp}

\end{document}